\documentstyle[12pt]{article}

\catcode`\@=11

\begin{document}

\baselineskip 24pt
\newcommand{\numero}{SWAT/27}
\newcommand{\titre}{STRONG EXTENDED TECHNICOLOUR INTERACTIONS}
\newcommand{\titreb}{AND THE $Zb\bar{b}$  VERTEX}
\newcommand{\auteura}{Nick Evans}
\newcommand{\addressa}{ }
\newcommand{\auteurc}{D.A. Ross }
\newcommand{\beq}{\begin{equation}}
\newcommand{\eeq}{\end{equation}}
\newcommand{\Fn}{\mbox{$F(p^2,\Sigma)$}}

\newcommand{\addressc}{Department Of Physics  \\   University   of
     Wales, Swansea\\ Singleton park, \\ Swansea \\ SA2 8PP \\ U.K. }
\newcommand{\abstrait}{LEP precision measurements of the $Zb\bar{b}$ vertex
coupling are sufficiently accurate to see the non-oblique corrections from
the heavy gauge boson responsible for the large top mass in standard extended
tecnicolour models. If the ETC couplings are strong the techni-fermion
condensate may be enhanced by several orders of magnitude allowing a top
mass of order 150GeV to be obtained for large values of the ETC breaking
scale sufficient to supress the ETC contribution to the $Zb{\bar b}$ vertex.
In this letter we investigate the degree of fine tuning of the ETC coupling
required in a realistic model. We conclude that the fine tuning need only be
of order $10\%$.
     }

\begin{titlepage}
\hfill \numero
\vspace{.5in}
\begin{center}
{\large{\bf \titre }}
{\large{\bf \titreb}}
\bigskip \\by\bigskip\\ \auteura \bigskip \\ \addressc \\

\renewcommand{\thefootnote}{ }
\vspace{.9 in}
{\bf Abstract}
\end{center}
\abstrait
\bigskip \\
\end{titlepage}

\def\id{\rlap{1}\hspace{0.15em}1}

Extended Technicolour (ETC) models \cite{ETC} attempt to explain the light
fermion mass spectra in terms of interactions with heavy techni-fermion
condensates mediated by the exchange of massive gauge bosons.
Recent direct searches for the top quark \cite{topmass} indicate that it has a
mass
in excess of 130GeV. If this mass is generated by standard ETC
dynamics
then the associated ETC gauge boson must be light ($M_{ETC}
\sim 1TeV$) . Calculation \cite{Chiv,Tern} shows that the non-oblique
corrections
to the $Zb\bar{b}$ vertex from such a gauge boson are sufficiently large
to be visible in current LEP precision measurements. A large top
mass can be obtained when the ETC gauge bosons are heavier
than that predicted in the standard ETC scenario if the ETC
interactions are themselves strong \cite{Appel}. In this letter we investigate
the degree of fine tuning of the ETC interactions required to
generate both a realistic top quark mass and a correction to the
$Zb\bar{b}$ vertex compatible with the LEP data.

In ETC models in which the top quark mass is generated by the
exchange of an $SU(3)_C \otimes SU(2)_L \otimes U(1)_Y$ neutral
ETC gauge boson,
with mass $M_{ETC}$ and coupling constant $g_{ETC}$, the
current to which it couples is constrained by the electroweak symmetry
to have the form

\beq \xi \bar{\Psi}^i_L \gamma^{\mu} T^i_L + \xi' \bar{t}_R^k
\gamma^{\mu} U^k_R  \eeq

\noindent where $\Psi_L = ( t,b)_L$, $T_L = (U,D)_L$ with U and D
techni-fermions, i and k are technicolour indices and $\xi$ and $\xi'$
are Clebsch-Gordon coefficients of order one
associated with the ETC gauge group.
The top mass is then given by

\beq m_t  \sim \xi \xi' \frac{g_{ETC}^2}{M_{ETC}^2}
<{\bar U}U>  \hspace{.25cm}\sim \frac{g_{ETC}^2}{M_{ETC}^2}4 \pi F_{\pi}^3 \eeq

\noindent where the techni-fermion condensate, $<\bar{U}U>$, has been estimated
by naive dimensional arguments \cite{Georgi} in terms of the techni-pion
decay constant,$F_{\pi}$,
that determines the W and Z gauge boson masses.

The interaction in Eqn(1) also gives rise to a correction to the tree level
$Zb\bar{b}$ vertex coupling \cite{Chiv}. In particular the ETC
interactions in Eqn(1) give rise to an effective operator (after
Fierz rearrangement)

\beq -\frac{1}{2} \xi^2 \frac{g_{ETC}^2}{M_{ETC}^2} (\bar{\Psi}_L
\gamma^{\mu}\tau^a\Psi_L)(\bar{T}_L \gamma_{\mu}\tau^a T_L) \eeq

At scales below the technicolour chiral symmetry breaking scale we
may replace the left handed techni-fermion current by the conserved
current associated with the $SU(2)_L$ symmetry in an effective
sigma model \cite{book}

\beq T_L \gamma_{\mu} \tau^a T_L \rightarrow \frac{1}{2} F_{\pi}^2
Tr(\Sigma^{\dagger} \tau^a iD_{\mu} \Sigma) \eeq

\noindent with $\Sigma = exp(2i\pi^a \tau^a/ F_{\pi})$ and $\pi^a$ the
techni-pions. There is then a tree level correction to the $Zb{\bar b}$
coupling in the effective Lagrangian

\beq \delta g_L^{ETC}  \sim - \frac{1}{2} \frac{g_{ETC}^2}{M_{ETC}^2}
F_{\pi}^2  \frac{e}{s_{\theta} c_{\theta}} I_3 \eeq

\noindent where $c_{\theta}$ and
$s_{\theta}$ are the cosine and sine of the weak mixing angle and $I_3$
the isospin of the external fermion. Substituting for $g_{ETC}^2/M^2_{ETC}$
from Eqn(2) we find

\beq \delta g_L^{ETC} \sim \frac{1}{4}  \frac{m_t}{4\pi F_{\pi}}
\frac{e}{s_{\theta}c_{\theta}} \eeq

This correction can be measured in the ratio of Z boson decay widths to
$b \bar{b}$ over that to all non$-b \bar{b}$ hadronic final states, for which
the
QCD corrections cancel in the light quark limit \cite{Chiv,Tern}.  We have

\beq \Delta_R = \frac{ \delta ( \Gamma_b / \Gamma_{h \neq b})}{ \Gamma_b /
\Gamma_{h \neq b}} \sim  \frac{2 \delta g_L g_L}{g_L^2 + g_R^2} \eeq

\noindent where $g_L = \frac{e}{s_{\theta}c_{\theta}} ( - \frac{1}{2} +
\frac{1}{3}s_{\theta}^2)$, $g_R = \frac{e}{s_{\theta}c_{\theta}} (\frac{1}{3}
s_{\theta}^2)$ . The contribution to $\delta g_L$  from longitudinal W
boson exchange and internal top quark loops has been calculated before
\cite{SMcorr}
and is of order $-0.7\% (-2.5\%)$ for $m_t = 100 (200)GeV$. The additional
contribution from ETC

\beq \Delta_R^{ETC} \sim -3.7\% \times \left( \frac{m_t}{100GeV} \right) \eeq

\noindent should be distinguishable by the LEP measurements with a
precision of $2\%$ \cite{Burgess} for $\Delta_R$, the result of Ref
\cite{Chiv}.

The dimensional arguments we have used above, however, are modified if
there are strong interactions at scales greater than the technicolour
confinement scale. Dynamical perturbation theory \cite{DPT} estimates
of the techni-pion
decay constant and hence the correction to the $Zb\bar{b}$ vertex given as
an integral equation over the full techni-fermion propagator  behave at high
momentum as

\beq    \sim  \int^{M^2_{ETC}} dk^2 \frac{\Sigma^2 (k)}{k^2} \eeq

\noindent whilst the techni-fermion condensate behaves at high momentum as

\beq \sim \int^{M_{ETC}^2} dk^2 \Sigma (k) \eeq

\noindent where $\Sigma(k)$ is the dynamical techni-fermion self energy.
Thus interactions that enhance the high momentum tail of
$\Sigma(k)$ will enhance the techni-fermion condensate and hence the light
fermion masses without enhancing either the W gauge boson masses or the
correction to the $Zb\bar{b}$ vertex.

This behaviour was first preposed
in Ref \cite{Walk} to motivate walking technicolour models in which the beta
function of the technicolour gauge group is assumed to be close to zero
effectively increasing the chiral symmetry breaking force at high momentum
in comparison with the usual running scenario and hence enhancing the tail
of $\Sigma(k)$ and the condensate. To numerically study the effects of
such enhancement it is necessary to truncate the Swinger Dyson equations;
the usual truncation is to  the gap equation \cite{num1,num2}. It is worth
stressing
that whilst the gap
equation approximation  shows the generally expected behaviour of chiral
symmetry break down for values of the gauge coupling above some critical
value the numerical values obtained for the fermion self energy
are unreliable. In particular the enhancement effect of a walking coupling
are sensitive to the precise form of the running of the gauge coupling constant
outside the perturbative regime. Sensible ansatzs in the non perturbative
regime must be well behaved as the momenta flowing through the coupling
tend to zero but the walking effect is sensitive to the size of the coupling
at the perturbative running cut off and any increase in the coupling at
momentum below the same cut off. For example the authors in Refs \cite{num1}
 and \cite{num2}
use similar and equally plausible ansatzs for the walking coupling but
whereas Ref \cite{num1} sees an enhancement in the condensate of upto three
orders of
magnitude the results of Ref \cite{num2} show an enhancement of only one order
of
magnitude. An analysis of whether a walking technicolour coupling provides
sufficient enhancement to raise the ETC mass scale enough to supress
the $Zb\bar{b}$ vertex contribution has been performed in Ref \cite{Tern}.
Even with an ansatz for the running gauge coupling that allows a large
chiral symmetry breaking contribution from the perturbative running regime
the condensate enhancement is insufficient to supress the contribution to
$\Delta_R$ below that visible at LEP since the running time upto $M_{ETC}$
is so short.

The fermion condensate may also be enhanced by strong ETC interactions.
In addition to the  ETC interactions that mix techni-fermions and the light
fermions
in Eqn(1) there will be
ETC interactions between the techni-fermions broken at the scale
$M_{ETC}$ which, if the ETC scale is suitably high, can be represented by
the four Fermi interactions

\beq \frac{g^2_{U}}{M_{ETC}^2}
 \bar{T}_L
U_R \bar{U}_R T_L   +  \frac{g^2_{D}}{M_{ETC}^2}
\bar{T}_L D_R \bar{D}_R T_L \eeq

The chiral symmetry breaking contribution of these interaction has been
studied in the gap equation approximation \cite{NJL}; for $g^2<8\pi^2$
the interaction is
not sufficiently strong to drive chiral symmetry break down but as $g$
rises above the critical value ($g_c^2 =8 \pi^2$) chiral symmetry is broken
manifesting as a momentum independent contribution to the techni-fermion's
self energy. The self energy is typically of order $M_{ETC}$. To obtain values
of the
self energy below $M_{ETC}$ $g/g_c$ must be fine tuned to the same
degree as the required ratio $\Sigma/M_{ETC}$.

We consider theories in which the techni-fermion's chiral symmetry is broken
by a strongly interacting technicolour group in the presence of ETC
interactions with couplings close to their critical values \cite{Appel}.
The strength of the ETC interactions may also be the result of enhancement
by additional interactions giving rise to operators of the form in Eqn(11).
The gap
equation for the techni-fermion's self energy in Landau gauge is then

\beq  \begin{array}{ccc}
\Sigma(p) & = & \frac{3 C(R)}{4 \pi} \int^{M_{ETC}^2}_0 \alpha(Max(k^2,p^2))
\frac{k^2 dk^2}{Max(k^2,p^2)} \frac{\Sigma(k)}{k^2+\Sigma^2(k)} \\
&&\\
&&+ \frac{g^2}{8\pi^2M_{ETC}^2} \int^{M_{ETC}^2}_0 k^2 dk^2
\frac{\Sigma(k)}{k^2+\Sigma^2(k)} \end{array}  \eeq

\noindent where C(R) is the casimir operator of the fermion's representation
of the gauge group. The behaviour of the solution to the gap equation
as a function of g is shown in Fig 1 for an SU(3) technicolour group with
techni-fermions in the fundamental representation and for $M_{ETC} = 10 TeV$.
We allow the technicolour
coupling to run according to the one loop result

\beq  \begin{array}{cccc}
\alpha (q) & = & \frac{1.5}{1+ 1.5 \beta \ln(q/\Lambda_{TC})}
& q \geq \Lambda_{TC} \\
&&&\\
& = & 1.5 & q< \Lambda_{TC} \end{array} \eeq

\noindent where the maximum value of the coupling $1.5 \sim 2 \alpha_C$
and $\alpha_C = \pi/3C(R)$ is the critical coupling value when the running is
neglected. We set $\beta = 1$, a typical value in a running theory.
For the interactions to correctly reproduce the observed
W and Z gauge boson masses we require that the techni-pion decay constant,
which we estimate by the Pagel Stokar formula \cite{Pagels}

\beq F^2_{\pi} = \frac{N_{TC}}{4\pi^2} \int^{M^2_{ETC}}_0 k^2 dk^2
\frac{\Sigma^2 - k^2\Sigma\Sigma'/2}{(k^2+\Sigma^2)^2} \eeq

\noindent gives $F_{\pi} = 123GeV$, a typical value in one family
technicolour models.

We observe from Fig 1 that the strong ETC interactions act to enhance the
chiral symmetry breaking from the technicolour interactions even for
values of the ETC coupling well below the critical value. The techni-fermion
self energy is enhanced only at high momenta since the value of $\Sigma(0)$
is fixed (up to a factor of two) by requiring the Pagel Stokar formula to
give the correct W and
Z gauge boson masses. For large values of the ETC coupling the
technicolour confinement scale $\Lambda_{TC}$ falls to a few hundred GeV
presumably placing an upper limit on the value of the ETC coupling.
As we have observed above the enhancement of
the high momentum behaviour of $\Sigma$ will directly enhance the
techni-fermion
condensate and hence the light fermion masses.

To investigate the degree of fine tuning required to generate a top mass
${\cal O}(150GeV)$ and a value of $\Delta_R$ within experimental bounds
we solve the gap equation in Eqn(12). We estimate the top mass by

\beq m_t = \frac{g^2_{ETC}}{M^2_{ETC}} \frac{D(R)}{8\pi^2}
\int^{M_{ETC}^2}_0 k^2 dk^2 \frac{\Sigma(k)}{k^2+\Sigma(k)^2} \eeq

\noindent where D(R) is the dimension of the techni-fermion representation,
and $\Delta_R$ from Eqns (3) (4) and (7). We note that we
have not included the feedback effect
of the top quark propagator in the techni-quark or top quark gap equations;
including these additional contributions would lower the value of the effective
critical coupling but not change
the behaviour of the gap equation solutions or the indication of the degree of
fine tuning of the input couplings. Neglecting
these contributions simplifies the numerical analysis. In addition we
maintain the four Fermi approximation to the heavy gauge boson propagators
even when the scale $M_{ETC}$ is low and we might expect higher dimensional
operators to be significant since again it is the behaviour of the gap
equation solutions that are of interest.
In Fig 2 we show numerical
results for $\Delta_R$ against $M_{ETC}$ in an SU(3) technicolour theory
in which for each value of $M_{ETC}$ the value of $g_{ETC}$ has been
tuned to give a top mass of 150GeV.
For the ETC and standard model contributions to $\Delta_R$ to lie within
the $2\%$ experimental precision of LEP we require that $\Delta_R^{ETC}$
is less than a
$1\%$ effect and hence from Fig 2 we see that we require $M_{ETC} >
5TeV$. The results are insensitive to the number of technicolours,
$N_{TC}$, in this
approximation since if we maintain the maximum value of $\alpha \sim 2\alpha_C$
the $N_{TC}$ dependence in the gap equation, Eqn(12), approximately cancels.
The
$N_{TC}$ dependence in the Pagel Stokar formula, Eqn(14), requires that
$\Sigma(0)
\propto 1/N_{TC}$ for a given Z mass (achieved by suitably tuning
$\Lambda_{TC}$). This dependence then cancels against the $N_{TC}$ dependence
in
the top quark self energy, Eqn(15), leaving the ETC dynamics independent of
$N_{TC}$.

In Fig 3 we plot $m_t$ against the ETC coupling constant for
the two scales $M_{ETC} = 10TeV$ and $M_{ETC} = 100TeV$ from which an estimate
of the fine tuning can be obtained. When $M_{ETC} = 10TeV$ to change the
fermion mass
from $5GeV$ (the bottom quark mass) to $100GeV$ we must change $g/g_c$ by
$ \sim 40\%$ but then
by  a further $\sim 10\%$ to obtain $m_t = 150GeV$. This is tuning of order
$10\%$. When $M_{ETC} = 100TeV$
to change $m_t$
from $5GeV$ to $100GeV$ we must change $g/g_c$ by $30\%$ and then
by only a further $\sim 1\%$ percent to obtain $m_t = 150GeV$. This is
tuning of order $1\%$.

In a realistic ETC model the ETC gauge coupling will presumably unify with
the technicolour coupling at the ETC breaking scale. Consider for example
an $SU(4)_{ETC} \rightarrow SU(3)_{TC}$ model with coupling described by
Eqn(13); the ETC coupling at 10TeV will then lie between
$0.6g_{crit}$ and $1.2g_{crit}$ (where $g_{crit}^2 = 4\pi^2/3C(R)$ is the
critical
coupling of the ETC group
in the absence of a running coupling and hence an underestimate of the true
critical coupling) for values of the technicolour beta function between
1.5 and 1. At 100TeV it will lie between $0.4g_{crit}$ and $0.8g_{crit}$ for
the
same range of $\beta$. The tuning required with an ETC scale of
10TeV is of order the typical magnitude of the couplings at
that scale and is therefore not unnatural.

We also note that generating the large fermion masses by the enhancement
of the high momentum tail of the techni-fermions' self energy with strong
ETC interactions does not overly enhance the value of $\Sigma(0)$.
This mechanism is therefore a plausible method for generating the top
bottom mass splitting without generating a large splitting between the W and Z
gauge boson masses
and hence upsetting the stringent $\delta \rho$
(T) parameter bound.

In this letter we have shown in a numerical analysis of the gap equation
for the techni-fermion propagator that strong ETC interactions can enhance
the techni-fermion condensate sufficiently that a realistic value of the
top mass can be obtained for large values of the ETC scale. The ETC scale
can be raised until the non oblique corrections to the $Zb\bar{b}$ vertex
measured at LEP in the fractional shift in the ratio of Z hadronic widths
$\Gamma_b/\Gamma_{h \neq b}$ falls below the experimentally measured
precision. This result is obtained  at the expense of tuning the
ETC coupling to within $50\%$ of its critical value. To obtain a
top mass in excess of $100GeV$ the ETC coupling must be fine tuned to at least
${\cal O}(10\%)$.

\noindent {\Large \bf Acknowledgements}

The author would like to thank Warren Perkins and John Terning for
useful discussions and SERC for supporting this work.

\end{document}